\documentstyle[epsf]{mn}
\def\aboutless{\ifmmode {\mathbin{\lower 3pt\hbox
    {$\rlap{\raise 5pt\hbox{$\char'074$}}\mathchar"7218$}}}   
    \else {${\mathbin{\lower 3pt\hbox
    {$\rlap{\raise 5pt\hbox{$\char'074$}}\mathchar"7218$}}}   
    $\thinspace}\fi}
\def\aboutmore{\ifmmode {\mathbin{\lower 3pt\hbox
    {$\rlap{\raise 5pt\hbox{$\char'076$}}\mathchar"7218$}}}   
    \else {${\mathbin{\lower 3pt\hbox
    {$\rlap{\raise 5pt\hbox{$\char'076$}}\mathchar"7218$}}}   
     $\thinspace}\fi}
\def\muinv{\mu^{-1}}
\def\ellip{{\cal P}}

\title[Vacuum resonance in SGRs]
{Spectral Effects of the Vacuum Resonance\\
in Soft Gamma-Ray Repeaters}

\author[Bulik and Miller]
{Tomasz Bulik$^{1,2}$ and
M. Coleman Miller$^{1,3}$\\
$^1${University of Chicago, Department of Astronomy and Astrophysics,
5640 S. Ellis, Chicago,~IL 60637}\\
$^2${N. Copernicus Astronomical Center, Bartycka 18,
PL-00-716, Warsaw, Poland}\\
$^3${{\it Compton} GRO Fellow}}
\begin{document}

\label{firstpage}

\maketitle

\begin{abstract}
The association of all three soft gamma-ray repeaters (SGRs) with supernova
remnants has established that SGRs are young neutron stars, 
and has given us a starting point for detailed modeling.
One of the most popular classes of models involves strongly magnetised
neutron stars, with surface dipole fields $B\sim 10^{14}-10^{15}$ G.
In such strong magnetic fields, many otherwise negligible processes
can play an important role.  Here
we consider the effects of vacuum polarisation on Compton scattering.
Vacuum polarisation introduces a characteristic
density-dependent photon frequency
at which the normal modes of polarisation become nonorthogonal and the
mean free path of photons decreases sharply.  Our analytic results and
Monte Carlo simulations
of photon propagation through a magnetised plasma show that this effect
leads, under a wide range of physical conditions, to a broad 
absorption-like feature in the energy range $\sim$5 keV---40 keV.  We 
discuss this effect in light of the spectra from SGR 1806-20.
\end{abstract}


\section{Introduction}
 
   The first soft gamma-ray repeater (SGR) to be discovered,
SGR 0525-66 \cite{Mazets1,Mazets2},
was shown soon after its discovery to be positionally coincident with 
the N49 supernova
remnant in the Large Magellanic Cloud \cite{Evans}.  
More recently, the other two
SGRs (SGR 1806-20 and SGR 1900+14) have also been associated with
supernova remnants 
\cite{Kouveliotou1,Kouveliotou2,Hurley,Kulkarni1,Kulkarni2}.  
These associations have led to a consensus that
SGRs originate from young neutron stars.  

   The peak luminosities inferred for bursts from SGRs are typically
$10^{40}-10^{42}\,$erg s$^{-1}$ for $\sim 0.1-1$ seconds (excepting the
unusual March 5, 1979 event from SGR 0525--66, which had an inferred peak
luminosity of $\sim 2\times 10^{45}\,$erg s$^{-1}$).  These luminosities
are $\sim 10^2-10^4$ times the Eddington luminosity $L_E\equiv
1.3\times 10^{38}(M/M_\odot)\,$erg s$^{-1}$ at which the radial radiation 
force from Thomson scattering exactly balances the gravitational 
force on a fully ionised hydrogen plasma.  Nonetheless,
the spectrum from SGR 1806-20, which has burst more than 110 times and
is thus the best-studied of the SGRs, appears not to undergo the reddening
one would expect if the burst created an expanding optically thick
fireball \cite{Fenimore}.  Thus, despite the highly 
super-Eddington luminosities,
the plasma is probably confined near the surface of the neutron star.
Magnetic fields could confine the plasma, and perhaps suppress plasma
instabilities, with pulsar-like field strengths of $\sim 10^{12}\,$G
(Lamb 1982; Katz 1982, 1993, 1994).  

   There are lines of argument suggesting much greater field strengths,
on the order of $10^{14}-10^{15}\,$G.  As suggested in Thompson and
Duncan (1995), a $\sim 6\times 10^{14}\,$G field could, within the
$\sim 10^4$ year lifetime of the N49 supernova remnant, spin a neutron
star down to the 8 s period seen in the March~5 event.  Moreover, a 
superstrong magnetic field
may be able to account for the   energetics and time scales of SGRs 
(see Thompson \& Duncan 1995 for one model).  These models
are extremely complicated, and there are many aspects of microscopic
physics that are important only in very strong magnetic fields.  Here we 
concentrate on the effects of vacuum polarisation and, in particular,
on the spectral effects of the second vacuum frequency.

   In strong magnetic fields $B\aboutmore B_c$, where $B_c
\equiv {m_e^2c^3\over{\hbar e}}=4.414\times
10^{13}\,$G is the magnetic field at which the electron cyclotron 
energy equals the electron rest mass energy, vacuum polarisation
from virtual $e^+e^-$ pairs can become significant (see, {\it e.g.},
M\'esz\'aros 1992 for a discussion).  In fact, for strong enough fields
the vacuum contribution to the dielectric tensor can exceed the plasma
contribution \cite{Pavlov2,ventura79,Pavlov1}.  
This has effects on the normal modes and cross sections for photons.
Well below the electron cyclotron frequency $\omega_B\equiv eB/m_ec$, the 
polarisation normal
modes for photons in a very strong magnetic field are close to linear over 
most photon frequencies and incident angles.  

   At most photon frequencies $\omega\ll\omega_B$, the scattering cross
section depends strongly on the polarisation of the photon.  If the
electric field vector of the photon is in the plane formed by the 
magnetic field and the photon propagation direction $\hat{k}$, the photon
is in the
parallel mode and the scattering cross section is $\sigma_1\sim
\sigma_T$, where $\sigma_T=6.65\times 10^{-25}\,{\rm cm}^2$ is the Thomson
cross section.  If the electric field vector of the photon is perpendicular
to the ${\vec B}-{\hat k}$ plane, the photon is in the perpendicular mode and
$\sigma_2\sim(\omega/\omega_B)^2\sigma_T\ll \sigma_T$; this reduction
in cross section is understandable, because in a strong magnetic field
it is easier to oscillate an electron along the field than to oscillate
it across the field.  

There is, however, a small range of photon
frequencies in which the vacuum and plasma contributions to the dielectric
tensor are comparable to each other and the normal modes become strongly
nonorthogonal over a broad
range of angles.  This frequency, called the second vacuum frequency (the
first vacuum frequency is near the cyclotron frequency), depends on the
magnetic field $B$ and plasma number density $N$: $\omega_{c2}\sim N^{1/2}
B^{-1}$ for $B\ll B_c$ and
$\omega_{c2}\sim N^{1/2}B^{-1/2}$ for $B\gg B_c$.  Vacuum polarisation
does not change the polarisation-averaged cross section, but it does
change the normal modes and redistribute the total cross section
between them: when $\omega
\approx \omega_{c2}$, $\sigma_1
\approx \sigma_2\approx\sigma_T$.   At any 
particular density the cross section is unusually high for a narrow
range of frequencies (see Figure~1).
Conversely, if we assume that most of the flux is
transported in the lower cross section mode, for any photon frequency there
is a density at which the cross section is high.  Thus, the optical depth
seen by a photon is greater than it would be in the absence of the enhanced
cross section at $\omega_{c2}$.  Because the resonance frequency varies
with density, in an atmosphere of varying density the vacuum resonance has
an effect on the spectrum in a wide range of photon energies.

  Here we expand on results first presented in Miller \& Bulik (1996).
  In \S\ 2 we give the normal modes and cross sections for photons in
a strongly magnetised plasma, including the effects of a significant
electron-positron pair density and of magnetic fields $B>B_c$.
In \S\ 3 we give a qualitative derivation of the upper and lower bounds
to the vacuum resonance feature as a function of magnetic field, scale
height, and the fraction of positrons.
In \S\ 4 we describe our numerical method for following photons as
they scatter through the atmosphere.  We also give the results of these
simulations, and show that an absorption-like feature is present at
reasonable optical depths for any magnetic field $B\aboutmore B_c$.  We
neglect the effects of photon splitting;
at the energies $\hbar\omega\aboutless$ 50 keV of interest here the
probability of photon splitting within $\sim 10^6$ cm is small (see
Adler 1971). We likewise ignore proton cyclotron scattering, because
we focus on scale heights $l\aboutless$ 1000 cm at a constant magnetic
latitude, implying that proton cyclotron scattering will
yield extremely narrow features, $\Delta\omega\over \omega \ll 1\%$,
 which would be unresolvable
by current detectors.  Finally, 
in \S\ 5 we discuss how our results, combined with observations, can 
constrain physical parameters in SGRs.

\goodbreak
\section{Vacuum polarisation effects}

It is convenient to describe radiative transfer in a magnetised medium 
in terms of normal modes.  Including both the
electric permittivity tensor $\epsilon$ and the magnetic
permeability tensor $\mu$, the dispersion equation becomes
\begin{equation}
\vec k \times \left[ \muinv \left( \vec k \times \vec E\right)\right] +
{\omega\over c} \epsilon \vec E =0 \, ,
\label{dispersion}
\end{equation}
where $\vec E$ is the electric field.
A simple and elegant way
of solving equation~(\ref{dispersion}),
describing the normal modes, and calculating cross sections has been
introduced by Gnedin~\&~Pavlov~(1974).
Given a system of coordinates
with the magnetic field along the $z$-axis, and
the wave vector in the $yz$ plane at an angle $\theta$ to the magnetic
field, we define
\begin{equation}
\begin{array}{rcl}
n_I \equiv& 1 + {1\over 4} (\epsilon_{xx} + \epsilon_{yy}\cos^2\theta
+\epsilon_{zz}\sin^2\theta-\epsilon_{xz}\sin 2\theta\\[3mm]
 & -\muinv_{xx} -\muinv_{yy}\cos^2\theta-\muinv_{zz}\sin^2\theta +
\muinv_{xz}\sin 2\theta ) \; ,
\end{array}
\label{nI-def}
\end{equation}
\begin{equation}
\begin{array}{rcl}
n_L \equiv& {1\over 4}(\epsilon_{xx} - \epsilon_{yy}\cos^2\theta
-\epsilon_{zz}\sin^2\theta+\epsilon_{xz}\sin 2\theta\\[3mm]
 & +\muinv_{xx} -\muinv_{yy}\cos^2\theta-\muinv_{zz}\sin^2\theta +
\muinv_{xz}\sin 2\theta ) \; ,
\end{array}\label{nL-def}
\end{equation}
and
\begin{equation}
n_C \equiv {i\over 2} ({ \epsilon_{xy} \cos\theta 
+\epsilon_{xz}\sin\theta})\, .
\label{nC-def}
\end{equation}
where $\muinv_{ab}$ is the $ab$ component of the inverse
$\mu$ tensor.
The roots of the dispersion equation are
\begin{equation}
n_j = n_I \pm \sqrt{ n_L^2 + n_C^2}\, , \label{disp-root}
\end{equation}
where $j=1,2$ indicates which normal mode is selected.
The refraction coefficients are given by the real part of $n_j$, and
the total
absorption coefficients are proportional to
the imaginary part of $n_j$; $\xi_j= (2\omega/c){\rm Im}(n_j)$, and
the cross sections in the two modes are
\begin{equation}
\sigma_j =(2\omega/c N_e){\rm Im}(n_j)\, . \label{master-xsect}
\end{equation}
   The
polarisation of the normal modes can be described by the position angle
$\chi_j$ between the major axis of the polarisation ellipse and
the projection of the magnetic field $\vec B$ on the plane perpendicular to
the wave vector $\vec k$, and the ellipticity $\ellip$, whose modulus
is equal to the ratio of the minor axis to the major axis of the polarisation
ellipse, and whose sign determines the direction of rotation
of the electric field. The ellipticity and position angle are
\begin{eqnarray}
\ellip_j = {r_j -1\over r_j+1}\, , &
r_j\exp(2i\chi_j) = {\displaystyle 
  -n_C \pm \sqrt{ n_L^2 + n_C^2}\over \displaystyle n_L}\,.
\end{eqnarray}
The polarisation vectors can be described by a complex variable
$b$, or by two real parameters $q$ and $p$
\begin{equation}
b  \equiv q+ ip = {n_L\over n_C}\, . \label{qpdef}
\end{equation}
A detailed discussion of the polarisation of the normal modes
is given by Pavlov, Shibanov, \& Yakovlev (1980).

Using the cyclic coordinates
$e_\pm = 2^{-1/2}(e_x \pm i e_y )$, $e_0 =e_z$ (M\'esz\'aros 1992, p. 95),
the polarisation vectors are
\begin{equation}
\begin{array}{rcl}
e_\pm^j &=& {1\over\sqrt{2}} C_j e^{\mp i\phi}\left( K_j\cos\theta \pm
1\right)\\[3mm]
e_0^j &= & C_j K_j \sin\theta
\end{array}
\label{vectors}
\end{equation}
where
\begin{equation}
K_j = b \left[ 1 + (-1)^j \left(1 + b^{-2}\right)^{1/2}\right] 
\label{K-def}
\end{equation}
and $C_j = (1+ |K_j|^2)^{-1/2}$ is the normalisation. It should be noted
that in general $K_j$ is complex.

\goodbreak
\subsection{Vacuum  effects in the absence of matter}

In the presence of a strong magnetic field the vacuum acquires a non-zero
polarisability tensor due to the existence of virtual
electron positron pairs, so that 
the permittivity tensor $\epsilon_{ij}$ and
the magnetic permeability tensor $\mu_{ij}$ of the vacuum become different 
from unity.  Charge symmetry implies that both $\epsilon_{ij}$ and
$\mu_{ij}$ are diagonal in a vacuum.

For example, if $B\ll B_c$
and $\hbar\omega \ll mc^2$, so that photon splitting and the creation of
real pairs can be neglected,
the dielectric tensor and magnetic permeability tensor
are given by
\begin{eqnarray}
\epsilon_{ab}-1 &=& \left(\begin{array}{ccc}
                       -2a & 0 &0 \\
                        0 &-2a & 0 \\
                        0 & 0 & 5a \end{array}\right) \,  , \\
\muinv_{ab}-1 &=& \left(\begin{array}{ccc}
                       -2a & 0 &0 \\
                        0 &-2a & 0 \\
                        0 & 0 & -6a \end{array}\right) \, ,
\label{epsmu-lowB}
\end{eqnarray}
where $\displaystyle a\equiv{\alpha\over 45\pi}\left({B\over B_C}\right)^2$
and $\alpha$ is the
fine structure constant.  Inserting
equation~(\ref{epsmu-lowB}) into equations~(\ref{nI-def}), (\ref{nL-def}),
and (\ref{nC-def}) we obtain
\begin{eqnarray} 
n_I^{vac}&=& 1 + {11\over 4}a \sin^2\theta\, , \nonumber \\
n_L^{vac}& =&- {3\over 4} a \sin^2\theta \, ,  \label{nICL-vac} \\
n_C^{vac}& = &0 \, . \nonumber
\end{eqnarray}
The roots of the dispersion equation in a
vacuum are real, and the refraction indices of the normal modes are
\begin{eqnarray}
n_1^{vac}= n_I^{vac} +n_L^{vac} = 1+ 2 a \sin^2\theta
\label{n1-lowB}\, ,\\
n_2^{vac}= n_I^{vac} -n_L^{vac} = 1+   {7\over 2} a \sin^2\theta
\label{n2-lowB} \, .
\end{eqnarray}
Note that knowledge of the refraction indices
$n_1^{vac}$ and $n_2^{vac}$ suffices to determine the influence of vacuum
polarisation on the propagation of light in the presence of matter.

The refraction indices $n_{1,2}^{vac}$ have been calculated for
$\hbar\omega\ll m_ec^2$
and arbitrary magnetic field by Tsai \& Erber (1975),
\begin{eqnarray}
n_1^{vac} = 1 + {\alpha\over 4\pi}\eta_1(h) \sin^2\theta \, ,\\
n_2^{vac} = 1 + {\alpha\over 4\pi}\eta_2(h) \sin^2\theta \, ,
\end{eqnarray}
where $\displaystyle h\equiv{ B\over B_c}$, and
\begin{eqnarray}
\eta_1(h) &=&  8\ln\Gamma_1 (1+h) -4h\ln\Gamma(1+h)
- {2\over 3}\Psi(1+h) \nonumber \\
& &  -2h\ln h-2h^2+2h\ln(2\pi)+ {1\over 3h} 
+ {1\over3} -8L_1   \\
 \eta_2(h)&=& -4h\ln\Gamma (1+h) + 4h^2\Psi(1+h) +2h\ln h \nonumber \\
& & + 2h [\ln(2\pi) -1] -4h^2 +{2\over 3}\, .
\end{eqnarray}
In these expressions $\Psi(x) = {d\over dx} \ln\Gamma(x)$,
\begin{equation}
\ln\Gamma_1(x) = \int_0^x dt\, \ln\Gamma(t) + {x\over 2}(x-1-\ln(2\pi))\, ,
\end{equation}
and $L_1 = {1\over 3} + \int_0^1 dx\, \ln\Gamma_1(1+x)\approx 0.249$.
The asymptotic limits of the functions $\eta_{1,2}(h)$ have been
found by Tsai \& Erber (1975). In the low field limit, $B\ll B_c$,
\begin{eqnarray}
\eta_1(h)&\approx & {14\over 45} h^2 - {13\over 315} h^4  \\
\eta_2(h)&\approx &{8\over 45} h^2 - {379\over 5040} h^4 \, ,
\end{eqnarray}
and when $B\gg B_c$
\begin{eqnarray}
\eta_1(h)&\approx &{2\over 3} h + \left( {1\over 3} + {2\over 3}\gamma 
-8L_1\right) \\
\eta_2(h) &\approx &{2\over 3} - h^{-1} \ln(2h)\, .
\end{eqnarray}

Using equations~(\ref{n1-lowB}) and~(\ref{n2-lowB}) we
 can rewrite equations~(\ref{nICL-vac})
\begin{eqnarray}
n_I^{vac} &= &  {1\over 2} (n_1^{vac} +n_2^{vac})\nonumber \\
 & = &
1 + {\alpha\over 8\pi}\sin^2\theta
(\eta_1(h) +\eta_2(h)) \\ 
n_L^{vac} & =&  {1\over 2} (n_1^{vac} - n_2^{vac})= {\alpha\over
8\pi}\sin^2\theta
(\eta_1(h) -\eta_2(h))\\ 
n_C^{vac} & = & 0\; .
\end{eqnarray}
  The normal modes are polarised linearly
for all angles of propagation except for the degenerate case
of propagation along the direction of the field when they are
circular. Such a behaviour could also be found by considering the
symmetry of the problem.

\goodbreak
\subsection {Propagation of radiation in a
 system with matter present}

Consider now a system in which the plasma density is low enough that
the contribution of plasma to the dielectric tensor is $|\epsilon_{ij}^{p}-
1|\ll 1$,
and the magnetic field is small  so that  $|\epsilon_{ij}^{vac}-1|\ll 1$
and $|\muinv_{ij}-1|\ll 1$. This condition is satisfied for
$\displaystyle {B\over B_c} < 40$  
and $\omega_p^2 \ll \omega_B \omega$, which holds for
\begin{equation}
{\hbar\omega\over mc^2} \gg 1.6\times 10^{-8} \left(N_e\over 10^{24}
{\rm cm}^{-3}\right) \left(B_c\over B\right)\, .\end{equation}

Such a system can be described by the dielectric tensor
\begin {equation}
\epsilon_{ab} = \epsilon_{ab}^{vac} +\epsilon_{ab}^{p}- \delta_{ab}\, ,
\end{equation}
and the vacuum permeability tensor. The plasma dielectric tensor
can be expressed as $\epsilon_{ab} = \displaystyle\delta_{ab} -
\left(\omega_p^2\over \omega^2\right)
\Pi_{ab}$,  where $\omega_p = \displaystyle \left( {4\pi N e^2\over m}
\right)^{1/2} $ is the plasma frequency and $\Pi_{ab}$ is 
the plasma polarisation tensor.
The plasma polarisation  tensor is diagonal in the cyclic coordinates,
and for cold electron plasma   is given by
\begin{equation}
\Pi_{\alpha\alpha}= {\omega\over \omega + \alpha\omega_B -i \gamma_r}
= {\omega \over \omega_t +\alpha\omega_B} \, ,
\label{coldPi}
\end{equation}
where $\alpha=-1,0,+1$, and $\gamma_r= (2/3)(e^2/mc^3)\omega^2$ is
the radiative width, and we denote $\omega_t = \omega -i\gamma_r$.
When there is an admixture of pairs, such that
the positron fraction is $f\equiv n_{e^+}/(n_{e^-}+n_{e^+})$, the cyclic 
components of the polarisation tensor are modified
\begin{equation}
\Pi_{\alpha\alpha}^f = f \Pi_{\alpha\alpha} + (1-f) \Pi_{-\alpha-\alpha}\, .
\label{coldPif}
\end{equation}

Inserting equation~(\ref{coldPi}) into
equations~(\ref{nL-def}) and~(\ref{nC-def}), we obtain
\begin{eqnarray}
n_I &=& 1  + {1\over 4}{\omega_p^2\over \omega} \left[ 
(1+\cos^2\theta) {\omega_t\over \omega_t^2-\omega_b^2}
+ \sin^2\theta {1\over\omega_t}
\right] + \nonumber \\
& & + {\alpha\over 8\pi}\sin^2\theta (\eta_1(h)+\eta_2(h)) \, ,
\label{nI-sys}
\end{eqnarray}
\begin{equation}
n_L = -{\sin^2\theta\over 4} \left[  {\omega_p^2 \over \omega\omega_t}
{\omega_B^2\over \omega_t^2-\omega_B^2 } 
+{\alpha\over 2\pi} (\eta_1(h) - \eta_2(h))\right] \, ,
\label{nL-sys}
\end{equation}
and
\begin{equation}
n_C= -{1\over 2} {\omega_p^2 \over \omega} \left(\omega_B\over
\omega_t^2 - \omega_B^2\right)\cos\theta\, .
\label{nC-sys}
\end{equation}
When $f\ne 0$, the parameter $n_L$ is not changed, whereas
\begin{equation}
n_C =  -\left({1 - 2f \over 2}\right) {\omega_p^2 \over \omega^2} 
{\omega\omega_B\over \omega_t^2 - \omega_B^2}\cos\theta\, .
\end{equation}
Inserting equations~(\ref{nL-sys}) and~(\ref{nC-sys}) into 
equation~(\ref{qpdef}), and neglecting terms proportional 
to $\gamma_r^2$ we obtain
\begin{eqnarray}
q &=& {1\over 1-2f}{\sin^2\theta \over 2\cos\theta} {\omega_B\over \omega}
\times \nonumber \\
& &  \times\left[ 1 + {\alpha\over 2\pi} (\eta_1(h)-\eta_2(h))
{\omega^2\over\omega_p^2} \left( {\omega^2\over\omega_B^2} -1\right)
\right] \, , 
\label{q-sys}
\end{eqnarray}
and
\begin{equation}
p={1\over 1-2f}{\sin^2\theta \over 2\cos\theta} 
{\omega_B\gamma_r\over \omega^2}
\left[ 1 - {\alpha\over\pi} (\eta_1(h)-\eta_2(h)) {\omega^2\over\omega_p^2}
\right] \, .
\label{p-sys}
\end{equation}
The normal modes are linear and orthogonal for most frequencies and
angles, since $q^2\gg p^2$ for most of the parameter space. The absorption
coefficients in this regime can be obtained by inserting
equations~(\ref{nI-sys}), (\ref{nL-sys}),and (\ref{nC-sys}) 
into~(\ref{disp-root}),
and noting that the ratio $n_L\over n_C$ is large and real. Thus, we
obtain:
\begin{eqnarray}
\xi_1 & = & N \sigma_T \sin^2\theta   \label{mu1-lin} , \\
\xi_2 & = &  N \sigma_T \left({\omega\over\omega_B}\right)^2 
              \, . \label{mu2-lin}
\end{eqnarray}

The critical points are where $q(\omega)=0$.
Such points exist when
\begin{equation}
{\alpha\over 8 \pi} {\omega_B^2\over \omega_p^2} \left[\eta_1(h) -\eta_2(h)
\right] > 1\, .
\end{equation}
Generally, for a system consisting of  electron plasma  and magnetised vacuum
there exist two frequencies for which  $q(\omega)=0$, one very close to the
cyclotron frequency, and the other at $\omega\ll \omega_B$. In this
work we will be interested in the latter. At this point, called
the second vacuum frequency, the normal modes are circular 
 for most propagation angles.
This leads to an increase of the
absorption of the low cross section mode and a simultaneous
decrease of the absorption in the high cross section mode.
From equation~(\ref{q-sys}), the second vacuum frequency is
\begin{equation}
\omega_{c2} = \omega_p \left[ {\alpha\over 2\pi}
 ( \eta_1(h) - \eta_2(h) )\right ]^{-1/2}\, ,
 \label{vacfreq}
\end{equation}
with the limits
\begin{eqnarray}
\hbar\omega_{c2} &=& \hbar\omega_p \left(15\pi/\alpha\right)^{1/2}
\left(B_c/B\right)  \nonumber \\
&=& 30 N_{26}^{1/2}\left(B_c/B\right)\,{\rm keV}
 \end{eqnarray}
  for $B\ll B_c$    and 
\begin{eqnarray}
\hbar\omega_{c2} &=& \hbar\omega_p\left(3\pi/\alpha\right)^{1/2}
\left(B_c/B\right)^{1/2} \nonumber \\ 
&=& 13 N_{26}^{1/2}\left(B_c/B\right)^{1/2}
 \, {\rm keV}
\end{eqnarray}
 for $B\gg B_c$, where $N_{26}\equiv N/10^{26}$ cm$^{-3}$.
Polarisation of the normal modes at $\omega_{c2}$ is determined
by the parameter $p$. The modes are circular when
$p^2 < 1$ and linear when $p^2 > 1$. In the case when
$p=1$ and $q=0$  the two polarisation modes  coincide. 
Inserting equation~(\ref{vacfreq}) 
into~(\ref{p-sys}) we obtain the value of $p$ at the
vacuum frequency:
\begin{equation}
p(\omega_{c2},\theta) = {\alpha\over 1-2f}{\sin^2\theta\over 3\cos\theta}
{\hbar\omega_B \over mc^2}\, .
\end{equation}
We can define a critical angle $\theta_c$ as an angle for which
$p(\omega_{c2}, \theta_c)=1$. For the magnetic fields considered
here $\theta_c$ is very close to $\pi/2$; $\pi/2 - \theta_c \approx
(\alpha/3) (B/B_c)$ for electron plasma.  Again inserting
 equations~(\ref{nI-sys}), (\ref{nL-sys}), and (\ref{nC-sys}) 
into~(\ref{disp-root}),
and  noting that the ratio $n_L \over n_C$ is imaginary and small
for the angles $\theta<\theta_c$, we obtain
\begin{eqnarray}
\xi_{1,2} &=& N\sigma_T\left[ {1\over 2} \sin^2\theta + \left({\omega\over 
\omega_c} \right)^2 {1+\cos^2\theta \over 2} \right. \nonumber \\
 &\pm &   \left. \left({\omega\over \omega_c} \right)^3 {\cos\theta\over 2}
        \sqrt{1 - \left({\alpha\over 1-2f}{B\over B_c}\right)^2 {\sin^4\theta\over
9\cos^2\theta} }  
        \right]\, . \label{mu12-circ}
\end{eqnarray}
In the case of propagation orthogonal to the field, when
$\theta > \theta_c$, the normal modes become linear and 
the absorption coefficients are given by equations~(\ref{mu1-lin}) 
and~(\ref{mu2-lin}).
 
The width of the resonance is associated with the extent of the
interval where $q(\omega,\theta)<1 $. It can be estimated 
as the inverse of the derivative of $q$ at the vacuum frequency
\begin{equation}
\nu_{vac} =\left(\left. {d q\over d\omega}\right|_{\omega_{c2}}\right)^{-1} = (1-2f) {\cos\theta\over \sin^2\theta} {\omega_{c2}^2\over \omega_B}\, .
\label{szer}
\end{equation}
The vacuum frequency and the shape of the vacuum resonance
do not depend on temperature since thermal effects alter
the dielectric tensor weakly for $\omega\ll\omega_B$.
Unlike the cyclotron resonance, the vacuum resonance is
a result of {\em global} properties of the medium and does not
depend on any individual particle resonance.

\begin{figure*}
\epsfxsize=\textwidth
\epsfbox{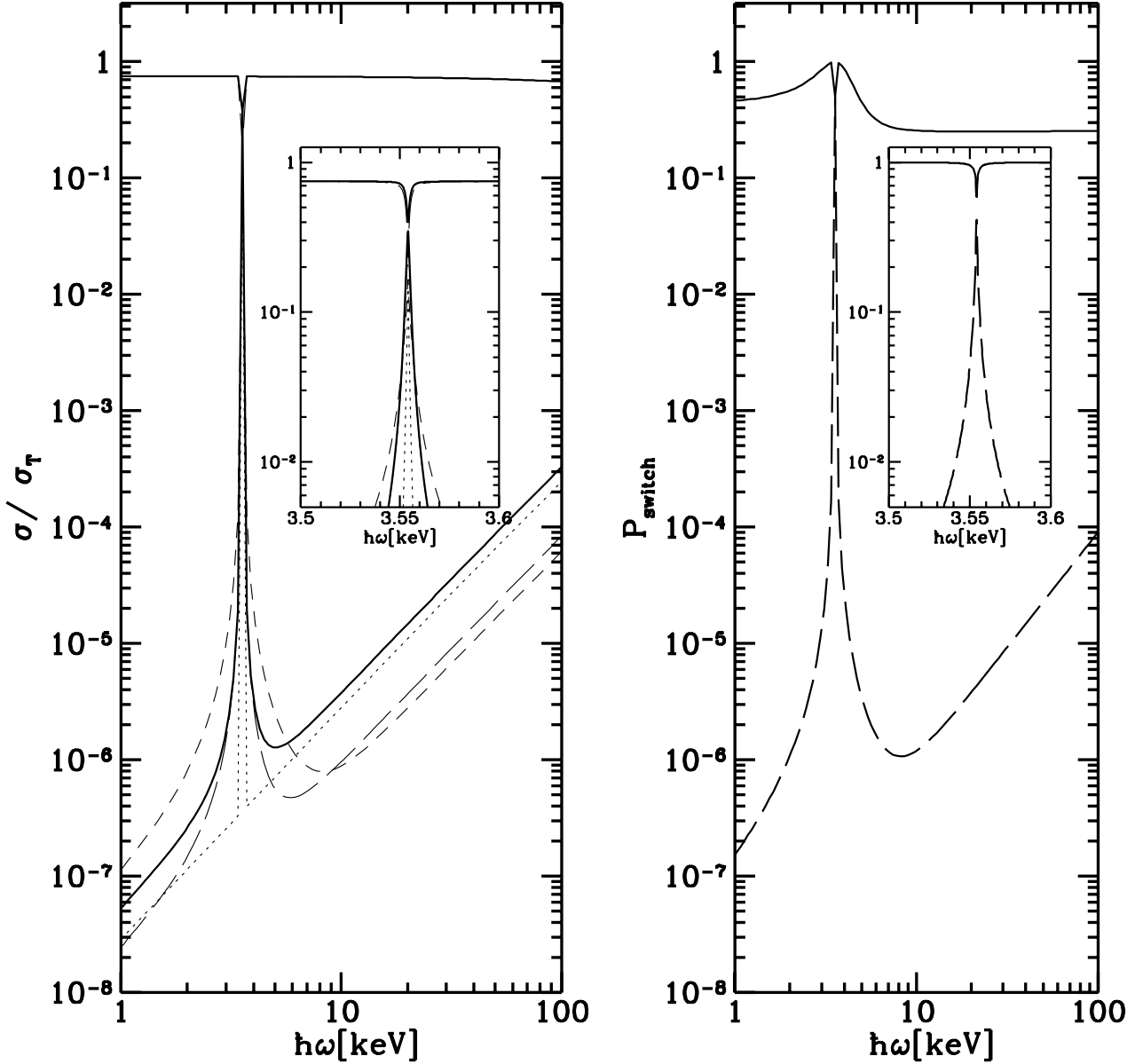}
\caption{Left panel: Scattering cross section, in 
units of the
Thomson cross section, versus photon energy.
The solid lines correspond to the total cross sections in the two
modes. The dotted line represents the   cross section  for the low mode to
 the low mode, the long dashed line represents the low mode to   high mode
cross section, and the short dashed  line represents the high mode to low
mode cross section; the high to high cross section is essentially
equal to the total high mode cross section. Note that the width of the 
resonance  in the low to high mode cross section exceeds that in   the low 
to low cross section. The inset shows a close up of the frequency range near the resonance. Right panel: Mode switching probabilities for the
low mode (solid line), and the high mode (dashed line).}
\end{figure*}

\goodbreak
\subsection{Scattering cross section}
   Let us now consider scattering in a hot, magnetised plasma.
A photon with energy $\omega$, wave vector
$k$ propagating at an angle
$\theta$ to the magnetic field $B$, and polarisation vector $e$
scatters on an electron with parallel
momentum $p$. After the scattering the photon has energy
$\omega'$, wave vector $k'$, polarisation vector $e'$,
and the electron momentum is $p'$.

The scattering cross  section is given by (M\'esz\'aros 1992, p. 131) 
\begin{equation}
d\sigma(\omega,\theta,p\mapsto\omega',\theta',p') =
r_0^2 {\omega'\over \omega} {m\over |k-k'|}
\left| \langle e|\Pi| e' \rangle\right|^2\, ,
\label{xsec00}
\end{equation}
where $r_0= e^2/m c^2$ is the classical electron radius.
After averaging equation~(\ref{xsec00}) over $\phi'$ we obtain
\begin{eqnarray}
d\sigma(\omega,\theta,p\mapsto\omega',\theta',p') &=& 
r_0^2  {\omega'\over \omega}{  m\over |k-k'|} \times \nonumber \\
 & & \times \sum_\alpha |\Pi_{\pm\alpha}|^2 |e_\alpha|^2 |e'_\alpha|^2\, ,
\label{cross-scat}
\end{eqnarray}
where we have the $+$ sign   for scattering on electrons and  $-$ for 
scattering on positrons. The scattering amplitudes, with 
first order relativistic corrections, are
\begin{eqnarray}
\Pi_+ &=& 1 - { \omega_B \over \omega' +\omega_B - pk' +
{{k'}^2/ 2} +i\gamma_r} \nonumber \\
\Pi_- &=& 1 + { \omega_B \over \omega - \omega_B - pk -
{k^2/ 2} + i\gamma_r}  ,\label{amplitudes} \\
\Pi_0 &=& 
 1+ { (p+k/2)(p-k'+k/2)\over \omega -pk -k^2/2 + i\gamma_r} \nonumber \\
& &  - { (p-k'/2)(p-k'+k/2)\over \omega' -pk' + {k'}^2/2 + i\gamma_r}
\, . \nonumber
\end{eqnarray}
 Assuming coherent
scattering, i.e. $k=k'$,  we average over the final states 
and use the polarisation normal modes  to  obtain the
total cross section from  one mode to another:
 \begin{equation}
\sigma_{ij} = \sigma_T \sum_\alpha |e_\alpha^i|^2 |\Pi_{\pm\alpha}|^2 
  A^j_\alpha  \, ,
\end{equation}
where $A^j_\alpha = {3\over 4} \int d \cos\theta |e^j_\alpha|^2 $.
In general the integrals $A^j\alpha$ are difficult to calculate analytically.
We can introduce the mode switching probability  for the two modes
\begin{equation}
p_{ij}= {\sigma_{ij}\over \sigma_{ij} + \sigma_{jj}}\, .
\end{equation}
We will find the cross section and the mode switching probabilities   
 in some limiting cases.

  When polarisation modes are linear for most of the
angles, e.g., outside the resonance, where polarisation is vacuum dominated,  we have $|b|=|n_L/n_c|\gg 1$, we find using equation~(\ref{K-def}) that
$K_1\rightarrow 2b$,
$K_2\rightarrow -1/2b$, 
$C_1\rightarrow 1/2b$, and $C_2\rightarrow 1$.  The components of the
two polarisation modes are then
\begin{equation}
e^1_\pm\approx {1\over{\sqrt{2}}}\cos\theta e^{\mp i\phi}\ ,\;
e^1_z\approx \sin\theta\; .
\end{equation}
and
\begin{equation}
e^2_\pm\approx \pm {1\over{\sqrt{2}}}e^{\mp i\phi}\ ,\;
e^2_z\approx 0\; ,
\end{equation}

The total scattering cross sections in the limit $u \equiv
 \left( {\omega_B\over \omega} \right)^2 \gg 1$ of photons
at initial incident angle $\theta$ to the magnetic field are
\begin{eqnarray}
\sigma_{1\rightarrow 1} &\approx & \sigma_T\sin^2\theta\; ,\label{sl11}
\\
\sigma_{1\rightarrow 2}&\approx &{3\over 4}\sigma_T{1\over u}
\cos^2\theta\; ,\label{sl12}
\\
\sigma_{2\rightarrow 1}&\approx &{1\over 4}\sigma_T {1\over u}\; ,
\label{sl21}
\\
\sigma_{2\rightarrow 2}&\approx &{3\over 4}\sigma_T {1\over u}\; .
\label{sl22}
\end{eqnarray}
Thus the  probability of  switching from the high cross section mode
to the low cross section mode is
\begin{equation}
p_{1\rightarrow 2} = {{3\over 4}\cos^2\theta \over u\sin^2\theta +{3\over 4}\cos^2\theta } \approx  {3\over 4}\left({\omega\over\omega_B}\right)^2 \cot^2\theta \, ,
\end{equation}
where the last approximation holds when 
$\theta \gg  \sqrt{3\over 2}{\omega\over\omega_B} $.
The probability of switching from the low cross section mode to the high cross section mode is
\begin{equation}
p_{2\rightarrow 1} ={1\over 4}\, .
\end{equation}

   In contrast, when $b\ll 1$, which holds for most of the propagation
angles in the resonance ($\theta<\theta_c$), the modes are circular, and
$K_1\rightarrow  1$, $K_2\rightarrow -1$,
$C_1\rightarrow 1/\sqrt{2}$, and $C_2\rightarrow 1/\sqrt{2}$. 
  The components of the two modes are then
\begin{equation}
e^1_\pm\approx \pm { 1\over 2}e^{\mp i\phi}(1\mp \cos\theta)\ ,\;
e^1_z\approx -{1\over \sqrt{2}}\sin\theta
\end{equation}
and
\begin{equation}
e^2_\pm\approx \pm {1\over 2}e^{\mp i\phi}(1\pm \cos\theta)\ ,\;
e^2_z\approx {1\over\sqrt{2}}
\sin\theta\; .
\end{equation}
The total scattering cross section in the limit $u\gg 1$ for any initial
or final polarisation state is then, in the limit $\sin\theta\gg u^{-1/2}$
\begin{equation}
\sigma_{i\rightarrow f}\approx {1\over 4} \sigma_T\sin^2\theta\; .
\label{sc-all}
\end{equation}
The mode switching probabilities are
\begin{equation}
p_{1\rightarrow 2} = p_{2\rightarrow 1} = {1\over 2}\, .
\end{equation}
 
We calculate the polarisation of the normal modes using the cold
plasma approximation, since the thermal effects are not important in the
range of energies far from the cyclotron resonance. We then use 
equation~(\ref{cross-scat}) and the scattering amplitudes~(\ref{amplitudes})
to find the cross section as a function of
the final angle. We present the  scattering cross section 
  as a function of photon energy
 in the left panel of Figure~1. In the right panel of Figure~1
  we show the mode switching probabilities for the
two modes. The insets show details  in the vicinity of the vacuum resonance.
Note that the low to high cross section mode
switching probability is enhanced around the vacuum resonance and reaches
nearly unity. Because of this, scattering in the low mode can
be dominated by mode switching, as it is in many of the results 
we describe  in \S\ 4.

   Absorption can also contribute significantly to the opacity over the
frequencies and densities of interest, depending on the composition of
the surface layers and magnetosphere.  If the primary atom has a high
atomic number, e.g., silicon or iron, then even when $kT=20$ keV the
atoms will not be completely ionised (atomic binding energies are increased
greatly by strong magnetic fields).  If the primary atom is hydrogen or
helium, virtually all atoms will be completely ionised and only scattering
and free-free absorption contribute to the opacity.  Here we assume that
only lighter elements are present, so that bound-free and bound-bound
absorption may be ignored.

   Given this assumption, scattering dominates the opacity over most of the
photosphere.  Free-free absorption can, however, be important for the
lower photon energies.  The free-free absorption cross
section in the low cross section mode is (see, e.g., M\'esz\'aros 1992)
\begin{equation}
\sigma_{abs}\approx 4\times 10^{-28}\left(B_c\over{B}\right)^{-2}N_{26}
T_8^{-1/2}\omega_{10}^{-1}\,{\rm cm}^2\; ,
\end{equation}
where $T_8\equiv T/10^8$ K and $\omega_{10}\equiv\hbar\omega/$10 keV.
When the photon energy $\hbar\omega<kT$, induced scattering and absorption
are important, and we include this by multiplying the total interaction
cross section by the factor $e^{\hbar\omega/kT}/(e^{\hbar\omega/kT}-1)$.

\subsection{Summary of the effects of vacuum polarisation}

We have shown above that vacuum polarisation affects strongly the
scattering cross section for the two modes. However, vacuum polarisation does not  introduce additional absorption. 
As can be seen from equations~(\ref{master-xsect}), and (\ref{disp-root}),
the polarisation averaged cross section depends only 
on the imaginary part of $n_I$. From equation~(\ref{nI-sys}),
  the vacuum contribution, represented
by  the third term in this equation, is real. 
Moreover, the imaginary part of $n_I$ is proportional to $N_e$.
 Therefore, from equation~(\ref{master-xsect}),
 the     polarisation averaged 
cross section     depends on neither plasma density  nor
 on   vacuum polarisation.
The main effect of
 vacuum polarisation is redistribution of the cross section between
the two modes. The effects described here will not be seen 
in  calculations that use polarisation averaged cross sections.

For example, the scattering cross section in the 
O-mode has no resonance at the cyclotron frequency when
the vacuum effects are neglected \cite{ventura79}.  
 Since radiative transfer takes 
place predominantly in the low cross section mode, neglecting vacuum polarisation would
inhibit formation of cyclotron lines in the polarisation averaged
spectra. 

Vacuum polarisation alters the mode switching probabilities,
which also influences   radiative transfer. In the case considered here, 
the increase of $p_{2\rightarrow 1}$ to almost unity in the wings of
the resonance also increases the number of scatterings that a photon undergoes
before leaving. This increases the chances of absorption, and 
enhances the effects of Comptonisation.

 \goodbreak
\section{Qualitative treatment of scattering with the
second vacuum frequency}

   From \S\ 2 it is clear that the scattering cross section depends 
strongly on the parameter
\begin{eqnarray}
b&\approx& {\sin^2\theta\over{2\cos\theta}}{\omega_B\over\omega}{1\over{(1-2f)}} \times
\nonumber \\
& & \times \left[1-{\alpha\over{2\pi}}\left(\omega\over\omega_p\right)^2
(\eta_1(h)-\eta_2(h))\right]\; .
\end{eqnarray}

    When $|b|\ll 1$ the cross section is $\sim\sigma_T$ in either polarisation
mode, whereas when $|b|\gg 1$ one mode has $\sigma\ll\sigma_T$.  Because
$\omega_{c2}\sim N^{1/2}$, as radiation propagates from regions of high
density to regions of low density the resonance frequency changes.  In
effect, the resonance acts as a sliding window of high opacity.
Qualitatively, we expect that the resonance will have an impact on the
flux at a given frequency if a) the optical depth through the resonance
is greater than unity, and b) the optical depth to the surface without the
resonance is less than unity.  These requirements give us, respectively,
the minimum and maximum photon energies at which the vacuum resonance
changes the spectrum.

   We can estimate the fractional width $\epsilon$ of the resonance by
arbitrarily setting the bound between $|b|\ll 1$ and $|b|\gg 1$ at
$|b|=1$ and
demanding that at $\omega=\omega_{c2}(1\pm\epsilon)$, $|b|=1$.  This gives
\begin{equation}
\epsilon={\omega_{c2}\over{\omega_B}}(1-2f){\cos\theta\over{\sin^2\theta}}
=0.02h^{-1}\omega_{10}{\cos\theta\over{ \sin^2\theta}}(1-2f)\; ,
\end{equation}
which is identical to equation~(\ref{szer}).
If $\epsilon\ll 1$ and the atmosphere is exponential with scale height 
$\ell$, then the thickness of the resonance is $\approx 4\epsilon\ell$.
Assuming that the resonance cross section is $\sigma\approx \sigma_T
\sin^2\theta$, if the magnetic field is parallel to the surface normal
then the optical depth through the resonance is
$\tau= N(4\epsilon\ell)\sigma_T\sin^2\theta/\cos\theta \,$:
\begin{eqnarray} 
 \tau & = & 6h\ell_1\omega_{10}^3(1-2f)\; ,\qquad    
B\ll B_c  \\
 \tau & = & 30\ell_1\omega_{10}^3(1-2f)\; ,\qquad  
B\gg B_c  ,
\end{eqnarray}
where $\ell_1\equiv (l/10$ cm).  This gives the 
lower bound to the feature.  To estimate the photon energy at which
the optical depth without the resonance is unity, we solve
\begin{equation}
n\sigma_T\left(\sin^2\theta\over{\cos\theta}\right)\left(\omega\over
{\omega_B}\right)^2\ell=1\; ,
\end{equation}
which gives $\hbar\omega\approx 30\ell_1^{-1/4}$ keV when $B\ll B_c$ and
$\hbar\omega\approx 20\ell_1^{-1/4}h^{1/4}$ keV when $B\gg B_c$.

   To summarise, when the magnetic field is parallel to the surface normal
then when $B\ll B_c$
this simple analytical treatment gives lower and upper
bounds $\omega_l$ and $\omega_h$ of
\begin{equation}
\hbar\omega_l\approx 6h^{-1/3}\ell_1^{-1/3}(1-2f)^{-1/3}\,
{\rm keV}\; ,
\end{equation}
\begin{equation}
\hbar\omega_h\approx 30\ell_1^{-1/4}(\sin\theta)^{-1}\,{\rm keV}\; ,
\end{equation}
whereas for $B\gg B_c$
\begin{equation}
\hbar\omega_l\approx 3\ell_1^{-1/3}(1-2f)^{-1/3}\,
{\rm keV}\; ,
\end{equation}
\begin{equation}
\hbar\omega_h\approx 20\ell_1^{-1/4}h^{1/4}(\sin\theta)^{-1}\,{\rm keV}\; .
\end{equation}
The scalings with magnetic field, scale height, and $f$ are confirmed by
our numerical results: 
typically $\hbar\omega_l\approx 5$ keV and $\hbar\omega_h\approx 40$ keV,
if the radiation is produced very deep in the atmosphere.

\begin{figure*}
\epsfxsize=\textwidth
\epsfbox{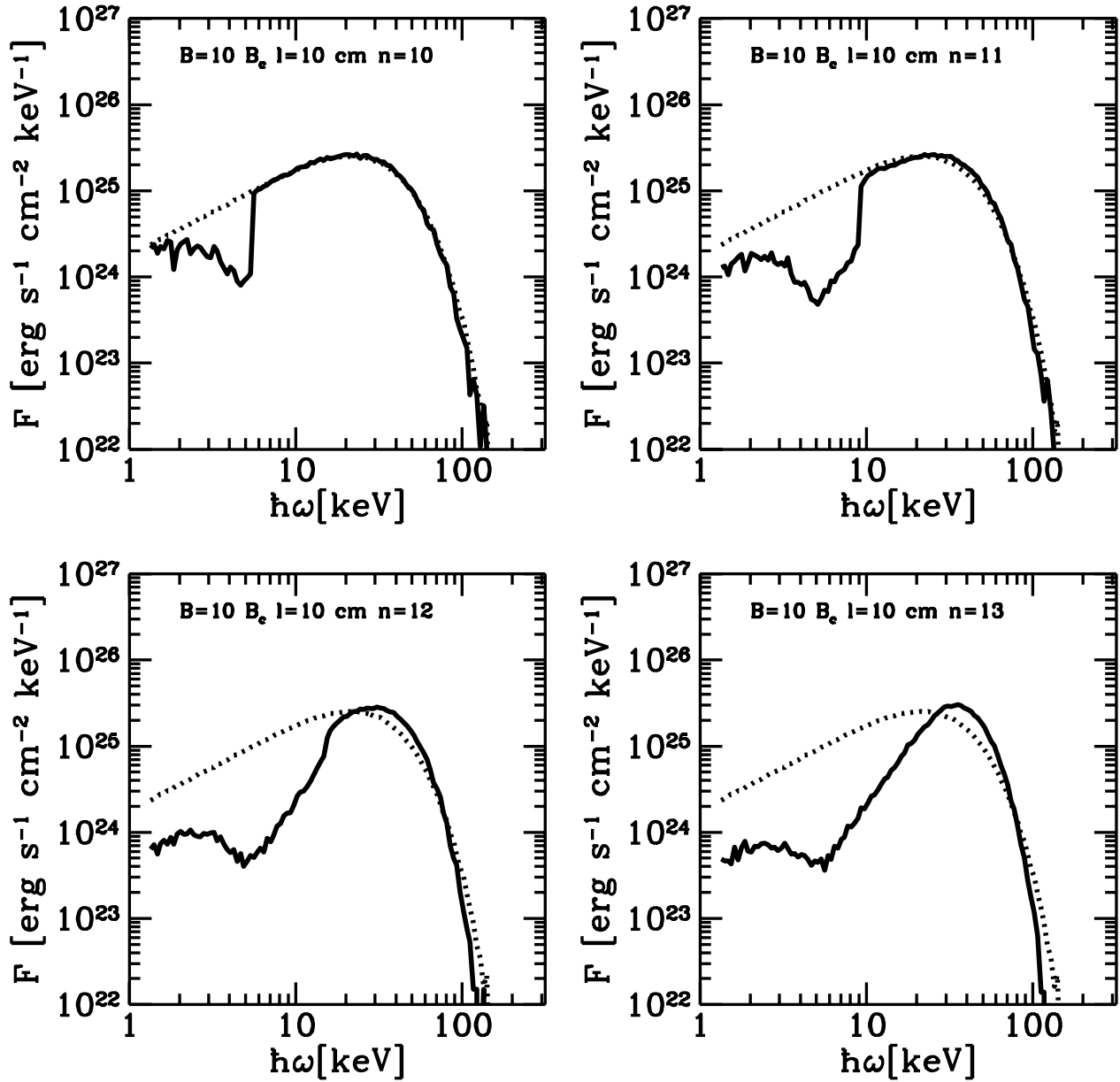}
\caption{ Model spectra, as seen at the stellar surface 
(solid lines), showing the effects of varying photon production depth.  The
input spectrum in each case is a 10 keV blackbody (dotted lines),
and the magnetic field is perpendicular to the surface.  The
magnetic field for all four panels is $B=10\,B_c$, the scale height is
$l$=10 cm, and the pair fraction is $f=0$.  In panels a, b, c, and d,
respectively, the photons are produced at 10, 11, 12, and 13 scale heights,
where the number density at the surface is $6\times 10^{21}$ cm$^{-3}$
(corresponding to $10^{-2}$ g cm$^{-3}$ for a fully ionised hydrogen plasma).
Note that the low-energy bound of the feature remains roughly constant,
whereas the high-energy bound increases with increasing production depth.}
\end{figure*}

\begin{figure*}
\epsfxsize=\textwidth
\epsfbox{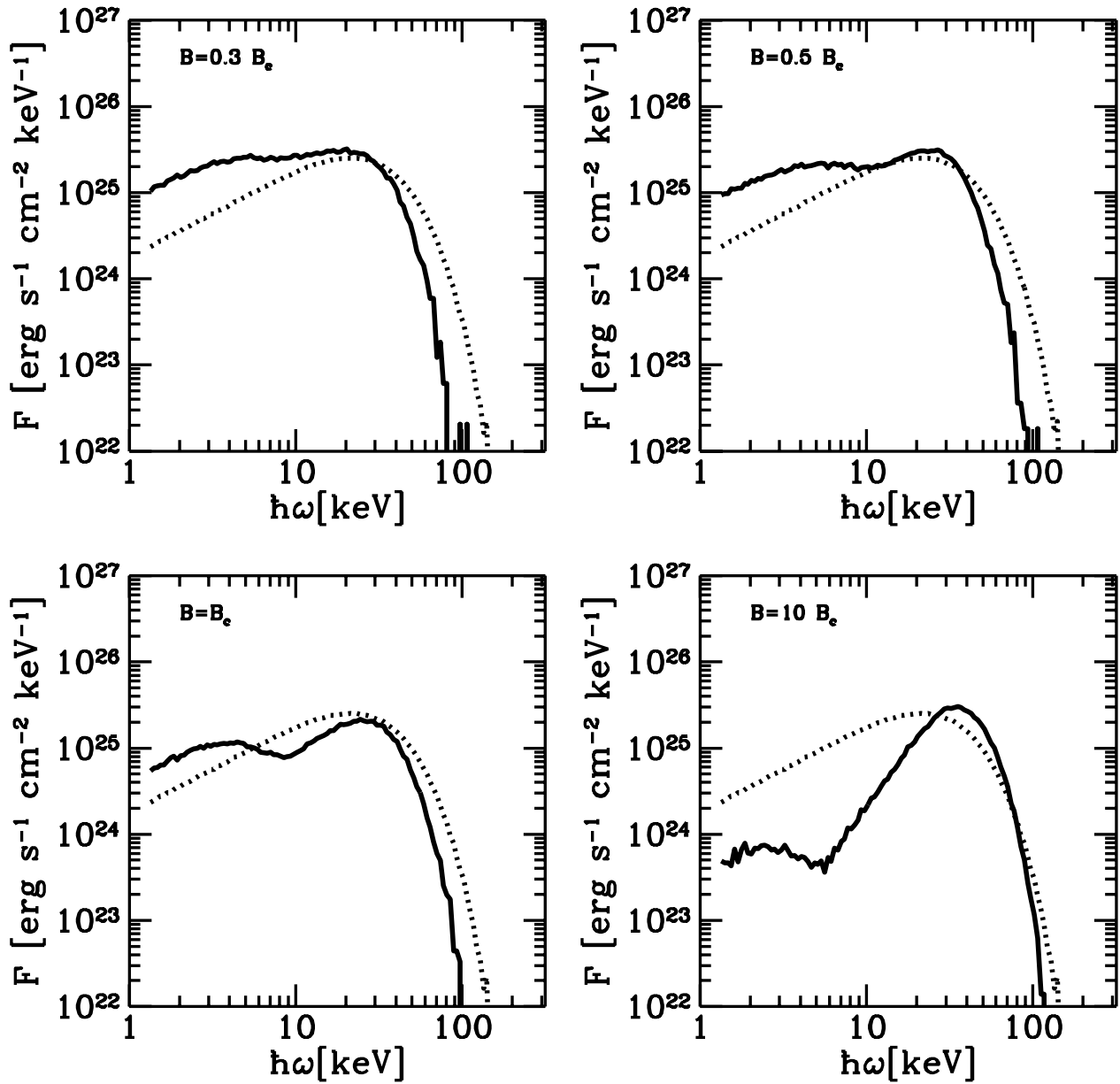}
\caption{ Model spectra, as seen at the stellar surface 
(solid lines), showing the effects of different magnetic fields.  As in
Figure~2, the input (dotted lines) is a 10 keV blackbody, 
the magnetic field is perpendicular to the surface, $l$=10 cm, 
and $f=0$, and here
the photons are assumed to be produced deep in the atmosphere.  Panels
a, b, c, and d, respectively, show spectra for magnetic fields of
$B=0.3\,B_c$, $B=0.5\,B_c$, $B=B_c$, and $B=10\,B_c$.  This figure shows
that the absorption-like feature will only be detectable for $B\aboutmore
B_c$.  }
\end{figure*}

\begin{figure*}
\epsfxsize=\textwidth
\epsfbox{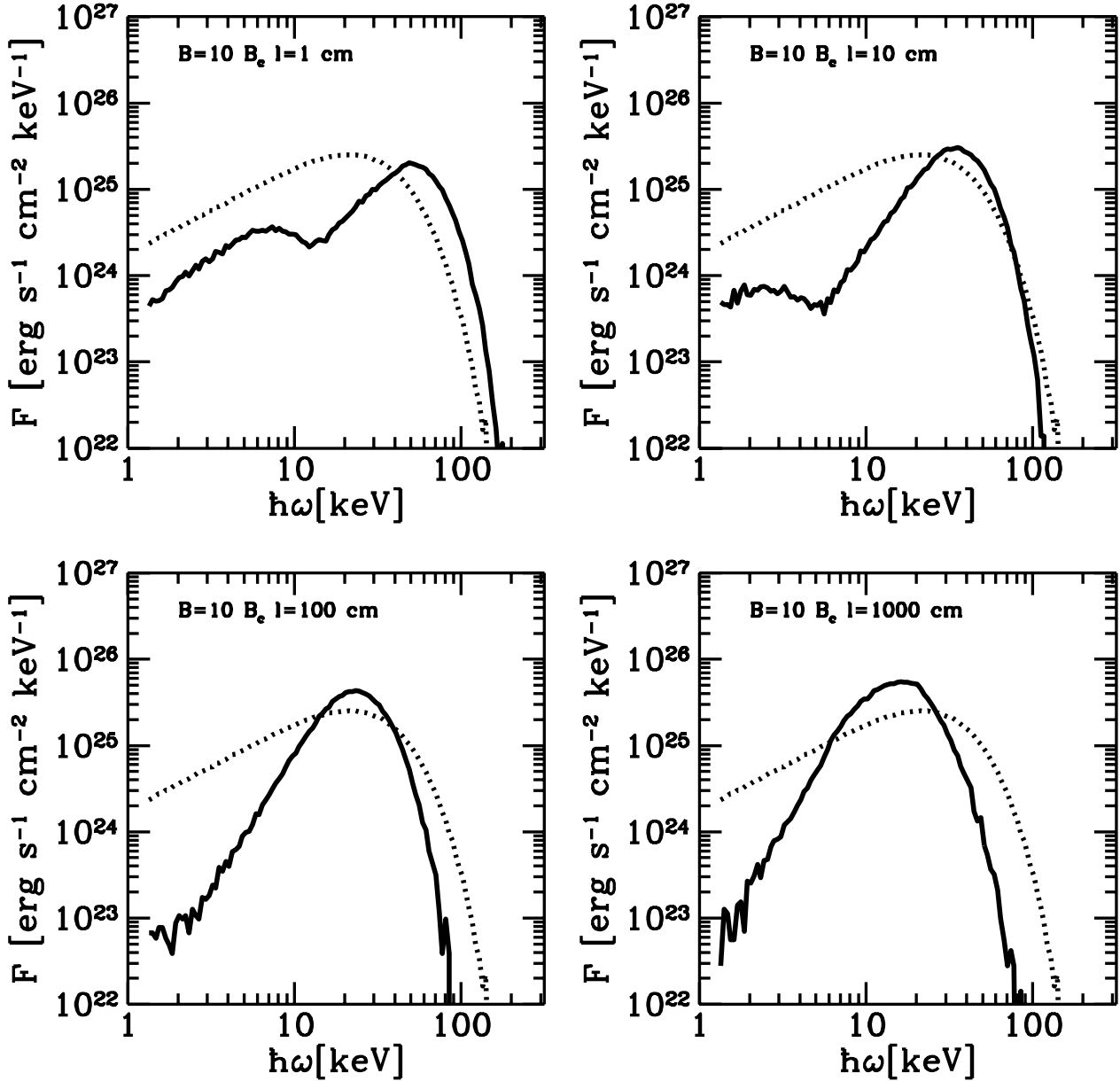}
\caption{  Model spectra, as seen at the stellar surface 
(solid lines), showing the effects of different scale heights.  Here
$B=10\,B_c$, $f=0$, the input (dotted lines) is a 10 keV blackbody 
produced deep in the atmosphere, and the magnetic field is perpendicular
to the surface.  Panels a, b, c, and d, respectively, show spectra
for scale heights of $l=1$ cm, $l=10$ cm, $l=100$ cm, and $l=1000$ cm.
Note that the vacuum feature moves to lower energies for larger scale
heights.  Thus, an instrument with a low-energy spectral cutoff of, e.g.,  
$\sim 5$ keV (such as ICE) might see only a low-energy dropoff during a
burst when the scale height is large, but could see the whole absorption-like
feature during the afterglow when the scale height has dropped.
  }
\end{figure*}

\begin{figure*}
\epsfxsize=\textwidth
\epsfbox{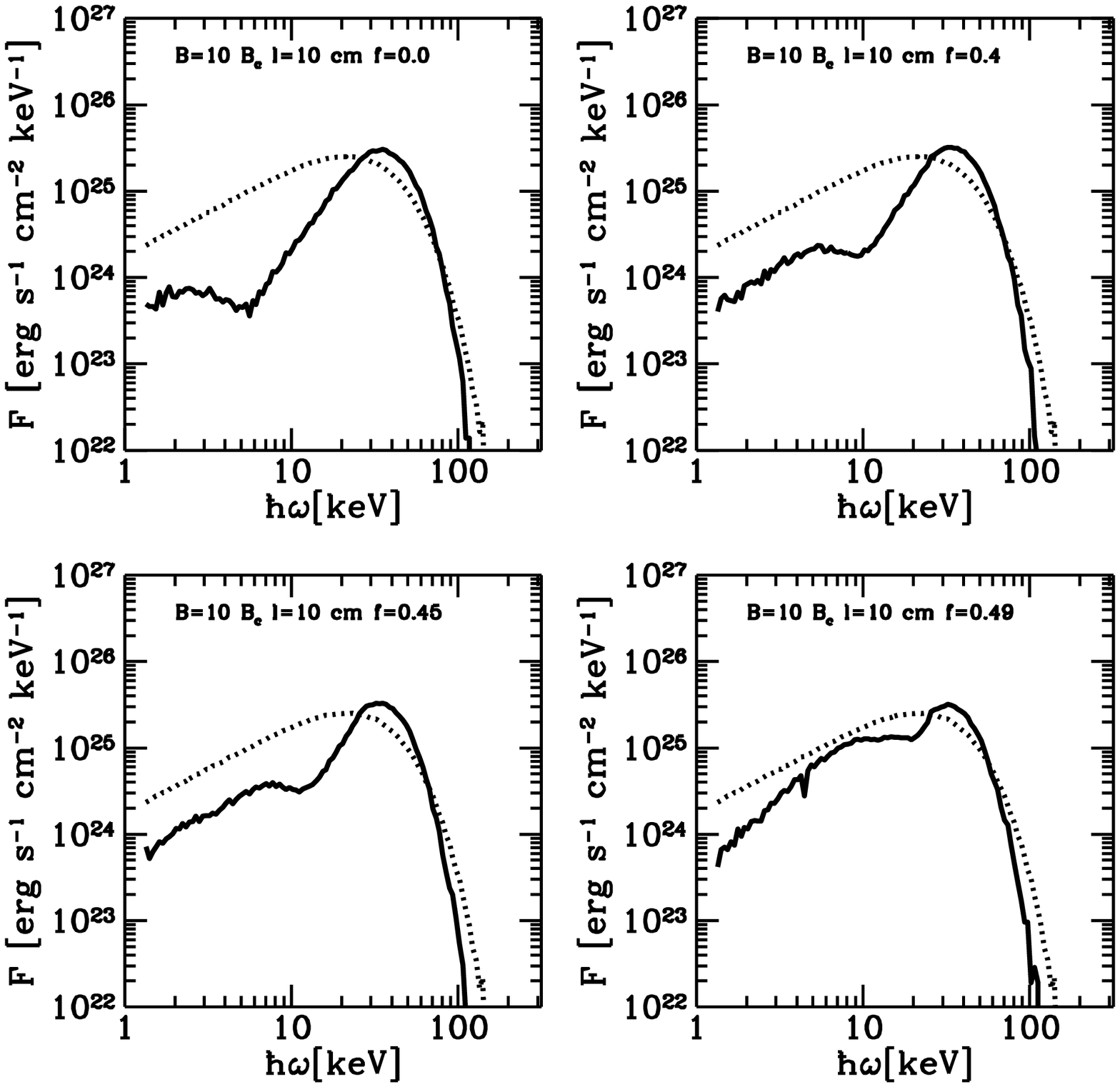}
\caption{   Model spectra, as seen at the stellar surface 
(solid lines), showing the effects of different pair densities.  In each
panel $B=10\,B_c$, $l=10$ cm, and the input (dotted lines) is a 10 keV 
blackbody produced deep in the atmosphere, and the magnetic field is
perpendicular to the surface.  In panels a, b, c, and d, respectively, the 
pair fraction (defined as the ratio of the number of positrons to the total
number of leptons) is $f=0$, $f=0.4$, $f=0.45$, and $f=0.49$.  As derived
in \S~3, the pair fraction affects only the low energy range of the
vacuum feature.
  }
\end{figure*}

\goodbreak
\section{Numerical method and results}

\subsection{Method}

   Following the procedure outlined in \S\ 2.3 to generate cross sections
and adding the corrections mentioned there for free-free absorption and 
induced Compton scattering, we use an adaptive
mesh to produce a table of physical depth as a function of optical depth 
for various values of the magnetic field and scale height.  The physical
depth depends on both the frequency and the direction of the photon,
and in particular for angles nearly perpendicular to the direction of the 
magnetic field the vacuum resonance feature is very sharp, and thus must be
integrated carefully.  We achieve sufficient accuracy with 130
logarithmically spaced zones in photon frequency from $\hbar\omega=
10^{-2.6}m_ec^2$ to $\hbar\omega=m_ec^2$, 50 linearly spaced
zones in angle, and optical depths from $10^{-1}$ to $10^4$, logarithmically
spaced.

   To produce a spectrum, we use a Monte Carlo code to pass
100,000 photons through the atmosphere, starting each photon at a
single layer deep in the atmosphere.
For each scattering
we generate an optical depth traveled, which combined with the direction
and with the table of physical depth versus optical depth gives the
new physical depth of the photon.  If the photon escapes from the 
atmosphere we store its energy and propagation direction.  Otherwise,
we use the differential cross section to select its new energy and
angle and determine if the photon switches polarisation modes, and repeat 
the process.  

   It is essential to include the effects of Comptonisation and
mode switching in these
simulations.  At temperatures of $\sim$10 keV the energy change
of a photon in a single scattering is much greater than the width of the
vacuum resonance.  Thus, in the $>$50\% of scatterings in the resonance
when the photon switches modes, the photon will be Comptonised
out of the resonance and may require several thousand scatterings
to return to the low cross section mode, during which its fractional
energy change can be of order unity and in the course of which the photon
is likely to be absorbed.  We account for the effects of mode
switching by checking at each scatter if the photon switches modes and,
if it does, simulating an absorption by drawing the new energy of the
photon from a thermal distribution.  We do not change the physical depth
of the photon in this case, because the mean free path in the
parallel mode is small enough that the net distance traveled in the parallel
mode is negligible.  To speed up the program we also
limit the total number of scatterings per photon to 1000; in the results 
reported here, between 97\% and 100\% of the photons escape in fewer than 
1000 scatterings.

\begin{figure}
\epsfxsize=0.5\textwidth
 \epsfbox{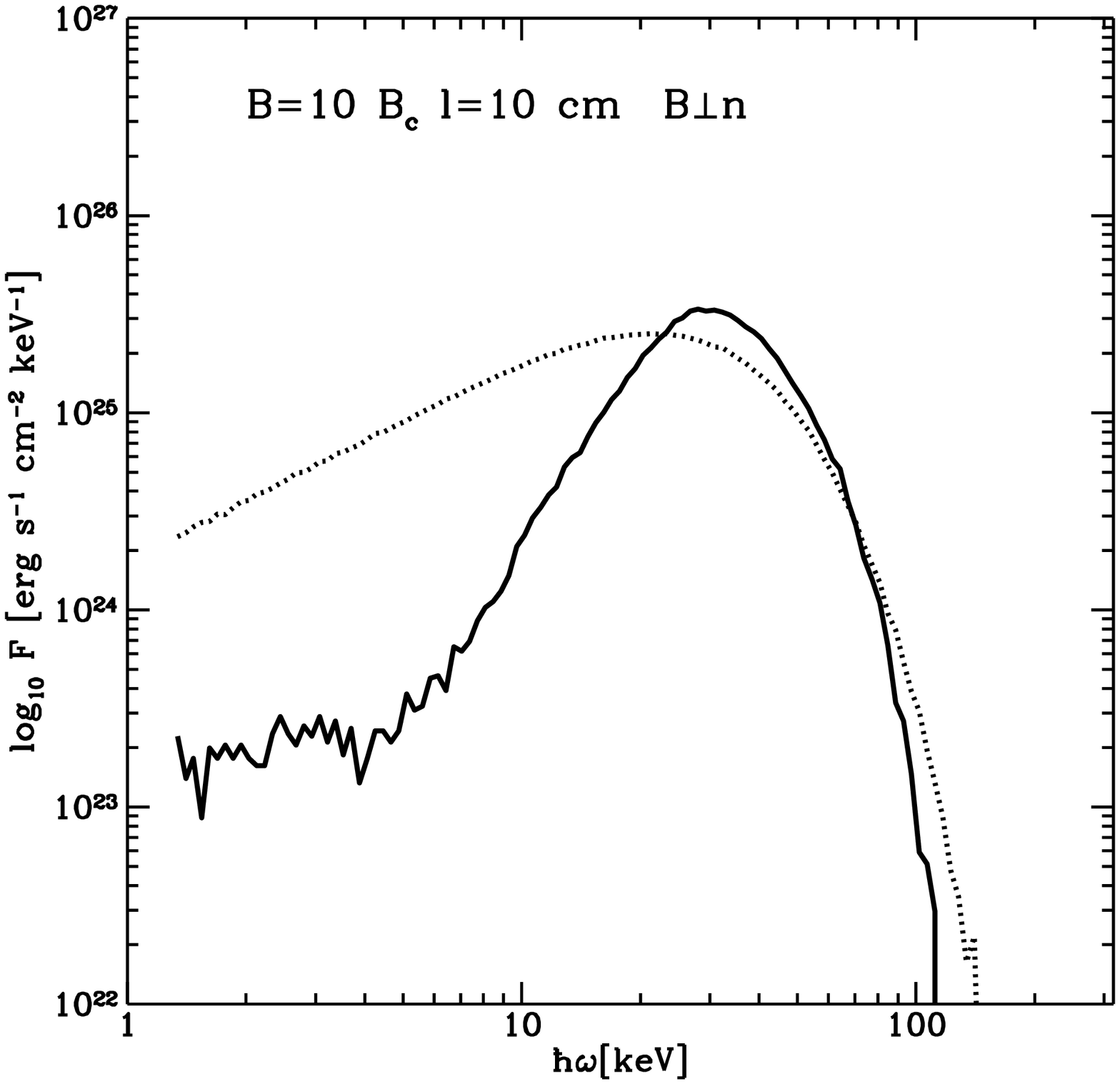} 
\caption{Model spectra, as seen at the stellar surface
(solid line), for a magnetic field in the surface of the star.  The
low-energy spectrum is flatter than when the field is perpendicular
to the surface.}
\end{figure}

\begin{figure}
\epsfxsize=0.5\textwidth
 \epsfbox{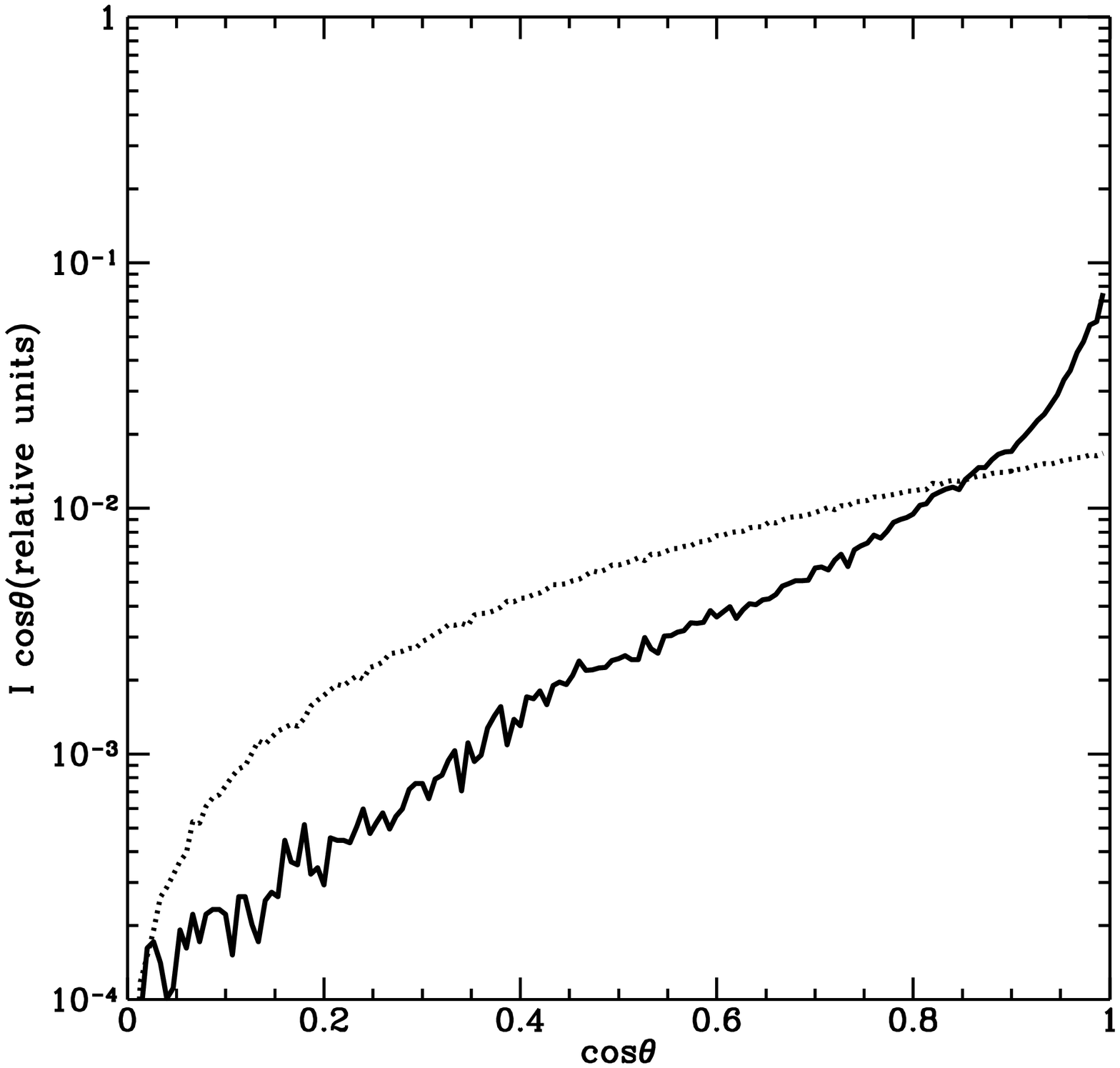} 
\caption{Angular distribution of emergent radiation
(solid line) for a magnetic field perpendicular to the surface, showing 
a strong peak along the surface normal ($\mu=1$).  For comparison,
we also show a simulation of the angular distribution for isotropic
scattering (dotted line). }
\end{figure}

\goodbreak
\subsection{Results}

   In this section we show spectra for several representative
combinations of physical parameters.  In every case the input spectrum is
a blackbody at 10 keV.  Figure~2 shows a sequence where the production depth
of the photons (in units of the scale height) changes but the magnetic field
is fixed at $B=10\, B_c$ and the scale height is 10 cm.
Note that the high-energy falloff
moves to higher energies with greater depth, whereas the low-energy edge
of the feature is roughly constant in energy, as predicted by our treatment
in \S~3.  Figure~3 shows spectra for different magnetic fields, with
$B=0.3$, 0.5, 1.0, and 10$\,B_c$.  In these the vacuum resonance feature is
only clearly visible for $B\aboutmore B_c$; at lower fields
the contrast in cross section in and out of the
vacuum resonance is not enough to create a significant feature.
In Figure~4 we isolate the effects of changing the scale height, as
$\ell$ ranges from 1~cm to 1000~cm.
From these we see that the larger the scale height, the broader
the feature, to the extent that the vacuum resonance may appear as only a 
falloff at low and high energies instead of an absorption-like feature.
In Figure~5 we concentrate on the effects of different pair densities.
In these graphs $B=10\,B_c$, the scale height is 10 cm, and the production
depth is 13 scale heights.  As the fraction $f$ of pairs (defined earlier)
increases from 0 to 0.49, the high-energy edge of the feature remains
fixed while the low-energy edge moves up, narrowing the trough and making
it shallower.  Figure~6 shows our last energy spectrum, in which the
magnetic field is in the surface instead of normal to it (as it was in
Figures~2-5), and we see that the low-energy part of the spectrum is
flatter than it is when the field is along the surface normal.

   In Figure~7 we show the angular distribution of emergent radiation,
integrated over all photon energies, for a magnetic field $B=10\, B_c$
along the surface normal, a scale height of 10 cm, and energy generated
deep in the atmosphere.  For comparison we also show the angular 
distribution of emergent radiation for isotropic scattering.  The
radiation is strongly peaked towards the normal; part of this comes
from geometric effects (see, e.g., Chandrasekhar 1960, \S 25), and
part comes from the angular dependences of scattering in a strong
magnetic field.  For a magnetic field in the surface, the emission also
depends on azimuthal angle, but qualitatively the angle-dependent
scattering and geometric effects partially cancel and the emergent
radiation is closer to isotropic than if only geometric effects operated.
We also note that, regardless of the orientation of the magnetic field,
the photon energy spectrum is essentially independent of viewing angle.

\goodbreak
\section{Discussion}

   Many of the spectra in the preceding section show prominent 
absorption-like features due to the vacuum resonance.  As
discussed in \S 4.1, these features
owe their existence in large part to the influence of Comptonisation
and mode switching, which imply that photons that interact in the
resonance undergo frequency diffusion and spread their energy over
the spectrum, creating a relative deficit at the original energy.
In contrast, if Comptonisation is neglected the photons undergo
primarily spatial diffusion, with minimal effects on the spectrum unless
the atmosphere is of uniform density (see, e.g., Ventura et al. 1979).
Thus, in treatments of this effect that do not
include Comptonisation (such as the Bezchastnov {\it et al.} 1996
radiative transfer approach) the vacuum resonance will be smeared
out by density gradients, whereas in our spectra the breadth and
visibility of the absorption-like feature are partially {\it because}
of the density gradients, as explained in \S 3.

   In the preceding section we examined the vacuum resonance features
produced for magnetic fields, scale heights, and pair fractions like
those expected in SGRs.  We now summarise the qualitative
effects of each parameter, and compare our theoretical spectra with the
spectra seen in SGR 1806-20.

   The physical quantity of greatest interest is the magnetic field, because
above $\sim B_c$ many exotic processes play an important role.  Our results
show that if a vacuum resonance feature is seen, the field must be stronger
than $\sim B_c$.  Since the location and strength of the feature depend
only weakly on $B$, observation of the feature would only place a lower bound 
on the magnetic field; this would imply that the field of the SGR was at
least as strong as the strongest magnetic field inferred for a pulsar,
$2\times 10^{13}$ G.

   If the density structure of the atmosphere can be represented as
exponential, then the scale height affects the qualitative nature of the
observed spectrum.  For large scale heights $\ell\aboutmore 1000$ cm, the
feature is manifested as a dropoff at $\sim 40-50$ keV, and the rise
at lower frequencies is below 5 keV.  If instead $\ell\aboutless 10$ cm,
both the low and high energy wings around the feature are in principle
visible above 5 keV.  These two regimes might be appropriate to a burst
(for large scale heights) and to afterglow (for small scale heights).
In particular, detailed time-resolved spectra should show this transition
as the luminosity drops.

   The depth of energy production affects the lower bound to the
vacuum resonance feature, and for a shallow enough production depth
(say, nine scale heights when $B=10\,B_c$) the feature is not present.
This implies that we may not expect to see such features in
accretion-powered pulsars, even if their fields are very strong;
typical stopping depths are $\aboutless$50 g cm$^{-2}$ (Miller,
Salpeter, \& Wasserman 1987), corresponding to less than nine scale
heights.

   The pair fraction $f$ does not affect the location of the upper
bound to the vacuum resonance feature, but does change the location 
of the lower bound.  If the plasma density is
dominated by pairs, the feature is narrowed and detectability requires
that the source of photons be deeper than when the electron-baryon
contribution dominates.  If the magnetic field and scale height are
known from other information, the centroid and width of the resonance
feature would then tell us the pair density and the depth where photons
were generated.

   How do the spectra of the bursts from SGR 1806--20 fit into this
picture?  None of the spectra have obvious spectral features that clearly
indicate a vacuum resonance.  There is an indication (Fenimore et al.
1994) that there is a low-energy falloff below $\sim 15$ keV; this
could be explained by a large scale height in our model, but many 
explanations not involving strong magnetic fields are equally viable.
The X-ray Timing Explorer, with its large area (3000 cm$^2$ at 3 keV)
and good spectral resolution (18\% at 6 keV) will, if it catches a
SGR burst, give us much-needed spectral information in the range
where features such as the vacuum resonance are expected to appear;
the High-Energy Transient Experiment may likewise provide useful
information, particularly with its Wide Field X-ray Monitor.

   In conclusion, in magnetic fields $B\aboutmore B_c$ and plasma densities
comparable to those expected in SGRs photon scattering is strongly
affected by the vacuum resonance.  Since scattering opacity is likely to
dominate over absorption opacity, this is an important effect which must be
included in calculations of radiative transfer through strongly magnetised
atmospheres.
\bigskip

{\bf Acknowledgements}
\medskip

We thank Carlo Graziani, George Pavlov, and Victor Bezchastnov for
valuable discussions.  This work was supported in part by NASA grant
NAG 5-2868 and, through the {\it Compton} Fellowship Program, by
NASA grant 5-2687.


\begin{thebibliography}{}
\def\bysame{\vrule height-1pt depth1.5pt width 65pt .\ }


\bibitem[Adler 1971]{Adler} Adler, S. L. 1971, Ann. Phys., 67, 599

 
\bibitem[Bezchastnov etal 96]{blabla}
Bezchastnov, V. G.,  Pavlov, G. G., Shibanov, Yu. A., \& 
Zavlin, V. E.  1996, Proceedings of the Third Huntsville
Workshop on Gamma-Ray Bursts, in press


\bibitem[Chandrasekhar 1960]{Chandra} Chandrasekhar, S. 1960, Radiative
Transfer (New York: Dover)

\bibitem[Evans et al. 1980]{Evans} Evans, W. D., et al. 1980, ApJ, 237, L7

\bibitem[Fenimore, Laros, \& Ulmer 1994]{Fenimore} Fenimore, E. E., 
Laros, J. G., \& Ulmer, A. 1994, ApJ, 432, 742

\bibitem[Gnedin \& Pavlov 1974]{Gnedin74}
Gnedin, Yu. N.,  \& Pavlov, G. G. 1974, JETP, 38, 903


\bibitem[Gnedin, Pavlov, \& Shibanov 1978]{Gnedin} Gnedin, Yu. N., 
Pavlov, G. G., \& Shibanov, Yu. A. 1978,
JETP Lett., 27, 305

\bibitem[Hurley et al. 1994]{Hurley} Hurley, K., Sommer, M., 
Kouveliotou, C., Fishman, G., Meegan, C.,
Cline, T., Boer, M., \& Niel, M. 1994, ApJ, 431, L31

 
\bibitem[Katz 1982]{Katz1} Katz, J. I. 1982, ApJ, 260, 371

\bibitem[Katz 1993]{Katz2} \bysame 1993, in Compton Gamma-Ray Observatory, 
ed. M. Friedlander, N. Gehrels, \& D. J. Macomb (New York: AIP), 1090

\bibitem[Katz 1994]{Katz3} \bysame 1994, ApJ, 422, 248

\bibitem[Kouveliotou et al. 1993]{Kouveliotou1} Kouveliotou, C. et al. 1993, 
Nature, 362, 728

\bibitem[Kouveliotou et al. 1994]{Kouveliotou2} Kouveliotou, C. et al. 1994, 
Nature, 368, 125

\bibitem[Kulkarni \& Frail 1993]{Kulkarni1} Kulkarni, S. R., \& Frail, 
D. A. 1993, Nature, 365, 33

\bibitem[Kulkarni et al.  1994]{Kulkarni2} Kulkarni, S. R., Frail, D. A., 
Kassim, N. E., Murakami, T., \& Vasisht, G. 1994, Nature, 368, 129

\bibitem[Lamb 1982]{Lamb} Lamb, D. Q. 1982, in Gamma-Ray Transients and 
Related Astrophysical Phenomena, ed. R. E. Ligenfelter, H. S. Hudson, \& 
D. M. Worrall (New York: AIP), 249

\bibitem[Mazets \& Golenetskii 1981]{Mazets1} Mazets, E. P., \& 
Golenetskii, S. V. 1981, Ap\&SS, 75, 47

\bibitem[Mazets et al. 1979]{Mazets2} Mazets, E. P., Golenetskii, S. V., 
Ilyinskii, V. N., Aptekar, R. L., \& Guryan, Y. A. 1979, Nature, 282, 587

\bibitem[M\'esz\'aros 1992]{Mesz} M\'esz\'aros, P. 1992, High-Energy 
Radiation from Magnetized Neutron Stars (Chicago: Univ. of Chicago Press)

\ 
\bibitem[mill and me]{ColeandMe}
Miller, M. C., \& Bulik, T. 1996, Proceedings of the Third Huntsville
Workshop on Gamma-Ray Bursts, in press


\bibitem[Miller, Salpeter, \& Wasserman 1987]{Miller} Miller, G. S.,
Salpeter, E. E., \& Wasserman, I. 1987, ApJ, 314, 215

\bibitem[Pavlov \& Gnedin 1984]{Pavlov1} Pavlov, G. G., \& 
Gnedin, Yu. N. 1984, Sov. Sci. Rev. E, 3, 197

\bibitem[Pavlov \& Shibanov 1979]{Pavlov2} Pavlov, G. G., \& 
Shibanov, Yu. A. 1979, ZhETF 76, 1457

\bibitem[Pavlov,  Shibanov, \&  Yakovlev 1980]{Pavlov3} Pavlov, G. G., 
Shibanov, Yu. A., \& Yakovlev, D. 1980, Ap\&SS, 73, 33

 
\bibitem[Thompson \& Duncan 1995]{Thompson} Thompson, C., \& Duncan, 
R. C. 1995, MNRAS, 275, 255

\bibitem[Tsai \& Erber 1974]{Tsai} Tsai, W., \& Erber, T. 1974, Phys. 
Rev., D10, 492

\bibitem[Ventura, Nagel, \&  M\'esz\'aros 1979]{ventura79}
Ventura, J., Nagel, W., \&  M\'esz\'aros, P. 1979, ApJ, 233, L125

\end{thebibliography}
\end{document}